\documentclass[twocolumn]{article}
\usepackage{epsfig}
\begin{document}
\markright{Submitted to {\em Nature}}

\begin{center}\large
\bf Time-Reversal Symmetry Breaking Superconductivity in Sr$_2$RuO$_4$
\end{center}
\begin{flushleft}\bf
G.~M.~Luke$^*$, Y.~Fudamoto$^*$, K.~M.~Kojima$^*$, M.~I.~Larkin$^*$,
J.~Merrin$^*$,
 B.~Nachumi$^*$,
Y.~J.~Uemura$^*$, Y.~Maeno$^\dagger$, Z.~Q.~Mao$^\dagger$,
 Y.~Mori$^\dagger$, H.~Nakamura$^\ddagger$, M.~Sigrist$^{\S}$ 
\end{flushleft}
\begin{flushleft}
$^*$Dept. of Physics, Columbia University, New York, NY 10027, U.S.A.\\
$^\dagger$Dept. of Physics, Kyoto University, Kyoto 606-8502, Japan.\\
$^\ddagger$Dept. of Material Science and
 Engineering, Kyoto University, Kyoto 606-8501, Japan.\\
$^{\S}$ Yukawa Instutute for Theoretical Physics, Kyoto University,
Kyoto 606-8502, Japan. 
\end{flushleft}

{\bf \boldmath In addition to its importance for
 existing and potential applications, superconductivity\cite{degennes} 
is one of the most interesting phenomena in condensed matter physics.
Although most superconducting materials are well-described 
in the context of the 
Bardeen Cooper and Schrieffer (BCS) theory\cite{bcs}, 
considerable effort has been devoted to the search for exotic systems
whose novel properties cannot be described by the BCS theory.
Conventional superconductors break only gauge symmetry  by selecting a definite
phase for the Cooper pair wavefunction;
a signature of an unconventional
  superconducting state is the breaking of additional 
symmetries\cite{ms91}.  Evidence for such broken 
symmetries  include anisotropic pairing
 (such as d-wave in the high-T$_c$ cuprates) and the presence of
multiple superconducting
phases (UPt$_3$ and superfluid $^3$He\cite{nobel}).  We have performed 
muon spin relaxation measurements of Sr$_2$RuO$_4$ and observe a 
spontaneous internal
 magnetic field appearing below T$_c$. Our measurements
indicate that the superconducting state in Sr$_2$RuO$_4$ is characterized
by broken time reversal symmetry which, when combined with symmetry
considerations indicate that its superconductivity  is
of p-wave (odd-parity) type, analagous to superfluid $^3$He. 
 Despite the structural similarity with
the high T$_c$ cuprates, the origin of the unconventional superconductivity
in Sr$_2$RuO$_4$ is fundamentally different in nature.}

Sr$_2$RuO$_4$, which is isostructural to 
the high-T$_c$ cuprate La$_{1.85}$Sr$_{0.15}$CuO$_4$,
 is to date the only known layered  perovskite superconductor which does
not contain copper.  
Although  first synthesized in
 the 50's,\cite{jjr59}
its superconductivity was only found in 1994\cite{ym94a};
 T$_c$'s of early samples were roughly
0.7~K but have increased to T$_c=1.5$~K in recent high
 quality single crystals\cite{apm98a}.  
%
%
%
%
Despite its low transition temperature, Sr$_2$RuO$_4$ is of great interest as
there is growing 
evidence for an unconventional superconducting state.
In this system, strong correlation effects enhance
the effective mass seen in quantum oscillation\cite{apm96}
 and Pauli spin susceptibility
measurements, in the same way as in $^3$He\cite{dv_book}.
Combining this feature with Sr$_2$RuO$_4$'s expected
tendency to display ferromagnetic spin fluctuations,
Rice and Sigrist\cite{tmr95}, and later Baskaran\cite{gb96}
argued that the pairing in
Sr$_2$RuO$_4$ could be of odd parity (spin triplet) type.

The strong suppression of the superconducting T$_c$ by even non-magnetic
 impurities suggests non-s-wave pairing\cite{apm98a}.
Specific heat\cite{sn97b} and NMR 1/T$_1$\cite{ki97a}
measurements indicate the presence of a large residual density
of states (RDOS) at low temperatures (well within the superconducting state);
in  high quality samples, this RDOS as T$\rightarrow0$
seems to approach half of the normal state value. 
Several authors\cite{ms96a,km96a} have proposed so-called non-unitary  
p-wave superconducting states for Sr$_2$RuO$_4$
to account for this RDOS  as well as the absence of a Hebel-Slichter
peak in NMR measurements\cite{ki97a}. 
A finite RDOS is not a unique signature of unconventional
superconductivity; for example it is observed in
  so-called 
gapless superconductors with isotropic s-wave pairing as a result of
 impurity scattering  (although this is unlikely in the specific
case of Sr$_2$RuO$_4$ where the finite RDOS apparently remains in the cleanest
samples).  It could also be explained with a
 multi-band hypothesis\cite{dfa97} where different gaps
are associated with  two types of bands. 
Therefore, further studies are required for a definitive 
determination of the pairing symmetry in Sr$_{2}$RuO$_{4}$.

One aspect of  the pairing symmetry, the breaking of time reversal symmetry
(TRS) can be probed directly.  If the superconducting state has a
degenerate representation (as is possible for some triplet superconducting
states) then TRS can be broken, whereas it cannot be broken for 
non-degenerate representations (the case for all singlet states).
When spin orbit coupling is small, the pair wave function
can be written as a product of the orbital part and the spin part.
Non-zero angular momentum of either the orbital or the spin
part could result in TRS breaking, although there are many such cases with
conserved TRS, such as d-wave states in the
high-T$_{c}$ cuprates.   TRS breaking is  also possible in the case
of strong spin orbit coupling.
In general, pairing states can be further classified in terms of 
gap functions attached to irreducible representations of a point group for
a given crystal lattice symmetry of the system\cite{ms91}.
TRS is broken for some of the unitary states and 
for all of the non-unitary states.  For example, the B phase
of $^{3}$He has a unitary state which conserves TRS, 
the A phase is unitary but breaks TRS, and so does the
non-unitary A$_{1}$ phase which is stable
 under an applied external magnetic field.  In contrast, conventional 
superconductors (including the BCS state) are unitary and conserve TRS.

For states with broken TRS, originating from either spin or 
orbital moments, a spontaneous internal magnetic field can appear 
below T$_{c}$.  One can expect a finite hyperfine
field at a magnetic probe for the case of spin moments, while
(for both the spin and orbital cases)
spontaneous supercurrents in the the vicinity of inhomogeneities in 
the order parameter  
would create a magnetic field near impurities, 
surfaces, and/or domain walls between the two degenerate 
superconducting phases\cite{ms91}.

Muon spin relaxation\cite{jhb94} is especially powerful for detecting
such effects of TRS breaking in exotic superconductors, since
internal magnetic fields as small as 0.1~G, 
corresponding to ordered moments $\sim0.01\;\mu_B$ are readily
detected in measurements without applying external magnetic fields.
Furthermore, the existence of a positively charged $\mu^{+}$ particle
could distort the electrostatic potential and perturb the superconducting
order parameter, 
inducing magnetic fields.  
$\mu$SR was previously applied to a search for a
spontaneous internal field predicted by an orbital state which breaks TRS 
in the ``anyon'' hypothesis proposed for high-T$_{c}$ cuprates.
No evidence for broken TRS was observed\cite{rfk90}  in that case.
However, a spontaneous internal field was observed by $\mu$SR 
in the so-called B phase of UPt$_{3}$\cite{gml93a}, and in 
(U,Th)Be$_{13}$\cite{rhh90}.  A number of authors have proposed states
(both unitary and non-unitary)
which break TRS for the B phase of UPt$_3$, consistent with this
 observation\cite{rhh_review}.

In a $\mu$SR experiment, 100~\% spin polarized positive muons are 
implanted one at a time into a specimen.  After quickly coming to rest
(at a typical depth of $\sim0.1$~mm), the 
muon spins evolve in the local magnetic environment.  The
muon subsequently decays ($\tau_\mu=2.2\;\mu$s), emitting a positron 
preferentially in the direction of the muon spin at the time of decay. 
 By accumulating time histograms of many ($10^7$) such positrons one
 may deduce the muon 
polarization function as a function of time after
implantation.  In a zero field (ZF-$\mu$SR) experiment, positron
detectors are positioned 180$^\circ$ apart, parallel and antiparallel
 to the initial
polarization direction.  Apart from normalization factors which depend on 
the experimental geometry and the properties of muon decay, the asymmetry
signal, which is defined as the difference  divided by the
sum of the count rates of the two opposing detectors, 
directly represents the muon 
polarization at the time of decay. 

 In the absence of magnetic order, the
 spin polarization is relaxed by static randomly oriented nuclear
 dipole moments and is well-described by the Kubo-Toyabe function:\cite{yju85}
\begin{equation}
{\cal P}_\mu = \frac{1}{3} +\frac{2}{3}\left(1-\Delta^2t^2\right)
\exp\left[-\frac{1}{2}\Delta^2t^2\right]
\end{equation}
where $\Delta/\gamma_\mu$ is the width of the local field distribution and
$\gamma_\mu=0.085\;\mu{\rm s}^{-1}{\rm G}^{-1}$ is the muon gyromagnetic ratio.
In the magnetically ordered state the $\mu$SR spectrum will exhibit precession
(where the frequency is proportional to 
 the ordered moment) if the internal field is
sufficiently uniform.  If the spontaneous field is weak, or there is a
broad distribution of local fields (as in, for example,
 a spin glass) then one observes
an increase in the relaxation of ${\cal P}_\mu$, where the increase in
 relaxation corresponds to an order parameter.

Two single crystals of Sr$_2$RuO$_4$ were grown at Kyoto University
 using a floating-zone method with an infrared image
furnace.  The feed rod,
containing 10\% excess Ru serving as flux, was melted in a 10\%-O2/90\%-Ar gas
mixture with a total pressure of 2 bar (Sample A) or 3 bar (Sample B).
 The crystal growth was performed
at speeds of 50 mm/hr (Sample A) and 40 mm/hr (Sample B). 
The superconducting T$_c$'s were 1.478~K and 1.453~K with widths of 
48~mK and 40~mK respectively, where  T$_c$ is defined as the sharp rise
in the dissipative component of the ac-susceptibility  and the width
  is its full width at half maximum.  Sample B was 
annealed in air at 1500 C for 72 h to reduce defects while sample A was
not annealed.
The fact that the two samples had high T$_c$'s (for Sr$_2$RuO$_4$) is indicative of their high quality and demonstrates that any impurities must
 be extremely dilute.\cite{apm98a} Sample A was 
cleaved and arranged in a mosaic 
so that the c-axis was normal to the sample's planar surface and
parallel to the initial muon polarization while  sample B was cut using a 
diamond saw such that the c-axis lay in the plane of the sample.  
Each specimen covered an area of roughly 1~cm$^2$.

We performed zero field (ZF-$\mu$SR) measurements on the M15 surface muon
channel at TRIUMF using an Oxford 400 top-loading dilution refrigerator.
  Sample A was mounted on 
a silver backing while sample B was mounted on high purity GaAs.  
Both  backing materials give a temperature-independent and
essentially non-relaxing $\mu$SR
 signal in zero field measurements.
Stray magnetic fields at the sample position were reduced to less than 0.1~G
in all directions.  Temperatures were measured with a calibrated carbon
resistor mounted on the mixing chamber as well as with a RuO$_2$ resistor
adjacent to the sample.

In Figure~1 we show ZF-$\mu$SR asymmetry 
spectra measured at T=2.1~K and T=0.02~K for sample B where 
${\cal P}_\mu\perp c$.  Spectra for sample A (${\cal P}_\mu\parallel c$)
are qualitatively similar.
From Fig.~1 we see that the relaxation rate is quite small both above and
below T$_c=1.45$~K,  but that
 there is greater relaxation at lower temperature.
This increased zero field
 relaxation could be caused by either (quasi-)static or
fluctuating magnetic fields.  To distinguish between these two possibilities,
we performed measurements in  a weak longitudinal field
B$_{{\rm LF}}=50$~G.  Relaxation due to static fields will be decoupled by
 the application of an external field greater than several times the
 characteristic internal field whereas decoupling of dynamic relaxation
 requires much greater applied fields\cite{jhb94}.  We see 
 in the
lower panel of Fig.~2 that the relaxation in
B$_{{\rm LF}}=50$~G  is the same above and below T$_c$,
which indicates that the  zero field relaxation must therefore
be  due to  spontaneous fields 
which are static on the $\mu$s timescale.

\begin{figure}
\begin{center}
\mbox{\epsfig{file=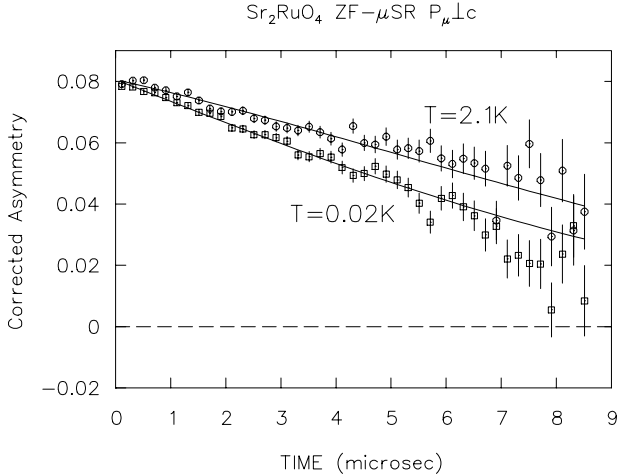,width=\columnwidth}}\end{center}
\caption[]{Zero field $\mu$SR spectra measured with 
${\cal P}_\mu\perp c$ in Sr$_2$RuO$_4$ at
T$=2.1$~K (circles) and T$=0.02$~K (squares).  Curves are fits to the product of Eqn.~1 and $\exp(-\Lambda t)$ as described in text.}
\end{figure}
 We tried several functional forms for the additional
relaxation below T$_c$;  fits to precessing (cosine) or gaussian forms were
essentially equivalent to each other, but were significantly worse than an 
exponential in fitting the data.
Fitting the spectra to the product of Eqn.~1 (using common values
of $\Delta= 0.02,\;0.06\;\mu{\rm s}^{-1}$ for samples A and B respectively,
 for all temperatures to account for the nuclear dipole fields) and
an exponential $\exp(-\Lambda t)$ to characterize the additional relaxation
due to the spontaneous magnetic field, we plot $\Lambda$ {\em vs.} T
 in Fig.~2 for samples A (${\cal P}_\mu\parallel$c) and B
(${\cal P}_\mu\perp$c).  Also shown are the superconducting T$_c$'s as determined by ac-susceptibility.  We see a clear increase in $\Lambda$ with an onset
temperature around T$_c=1.45$~K for both orientations indicating the presence of
a spontaneous magnetic field appearing within the superconducting state.
In principle, this spontaneous field could originate either from
a TRS breaking superconducting state or with a purely magnetic state
which coincidentally onsets near T$_c$.  Very recently, we performed
additional measurements on a third sample with a reduced T$_c=0.9$~K and
found that the spontaneous field onset at that reduced T$_c$. Thus, we 
conclude that since this field is so intimately connected with superconductivity,
 its existence  provides
 direct evidence for a time reversal symmetry breaking superconducting
state in Sr$_2$RuO$_4$.

\begin{figure}
\begin{center}\mbox{\epsfig{file=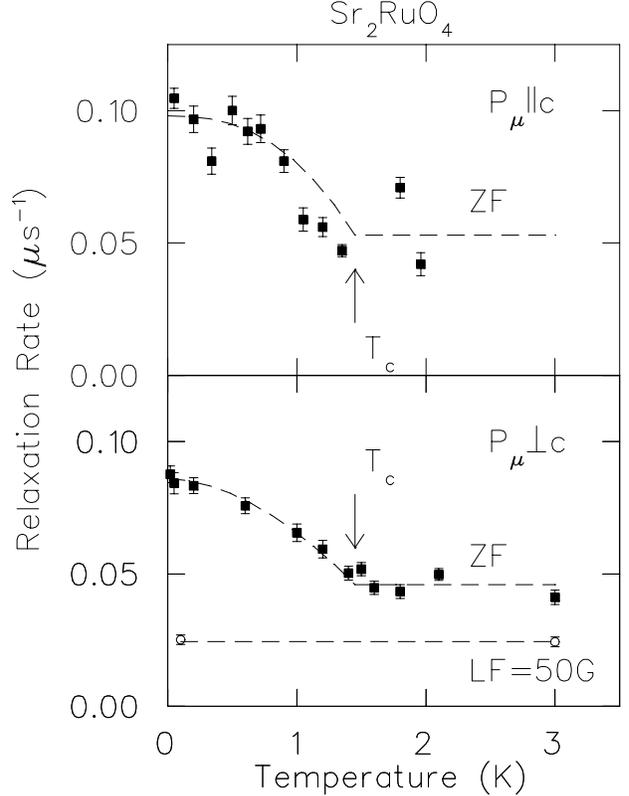,width=\columnwidth}}\end{center}
\caption[]{Zero field relaxation rate $\Lambda$ for the initial muon
spin polarization $\parallel c$ (top) and $\perp c$ (bottom).  T$_c$ from 
ac-susceptibility indicated by arrows.  Circles in bottom
 figure give relaxation rate in B$_{{\rm LF}}=50$~G~$\perp c$.
Curves are guides to the eye.}\end{figure}

The increased relaxation in the superconducting state is exponential
in character.
If there were a unique field at the muon site we would observe
a precession signal,
while
 a dense collection of randomly oriented
 field sources results in a Kubo-Toyabe relaxation 
function.  Both the precession and Kubo-Toyabe signals start with a 
gaussian form and in the limit of weak characteristic fields are 
essentially indistinguishable.   The exponential form which we 
observe in Sr$_2$RuO$_4$ indicates a broad
distribution of fields arising from a dilute distribution of
sources\cite{yju85}. This is consistent with the source of 
the internal field being
supercurrents associated with variations in the superconducting order 
parameter around dilute impurities and domain walls. 
We see enhanced relaxation in both samples A and B.  From this we can conclude
that the local field cannot lie purely along the c-axis (this would give no
relaxation from sample A).  However, any orientation of the local field
which has a significant component in the basal plane is consistent with our
data. 

The increase in the exponential relaxation below T$_c$ is about
0.04~$\mu {\rm s}^{-1}$ which corresponds to a characteristic field strength
$\Lambda/\gamma_\mu = 0.5$~G.  This is about the same as we observed in the B 
phase of UPt$_3$\cite{gml93a}
and is in line with theoretical predictions for that
material.  No theoretical estimates of the characteristic field strength
in Sr$_2$RuO$_4$ are yet available; however, we would expect them to 
be comparable to those in UPt$_3$ as the fields should arise from a similar
mechanism.

Several authors have
 considered the allowed  symmetries of pairing states for 
 the specific case of tetragonal symmetry appropriate for
Sr$_{2}$RuO$_{4}$\cite{tmr95,ms96a,km96a,ms91}.
Spin-singlet pairing leads exclusively to non-degenerate
states and the breakdown of TRS would, in general, require additional
 phase transitions admixing other pairing channels within the
 superconducting phase,
for which there is no experimental indication so far.  On the other hand,
one can show that in the spin-triplet pairing channel, TRS breaking 
states  appear naturally at the onset
 of superconductivity,
consistent with our experiment.  Based on these symmetry arguments we
 conclude that the present experiment provides strong evidence for Cooper
 pairing with spin-triplet (p-wave) symmetry: a superconducting 
analog of the A or A$_1$ phases of superfluid $^3$He.  
    The distinction between
 unitary and non-unitary states in Sr$_2$RuO$_4$,
however, cannot be done with the present results and has to wait for
 further studies by other means.

\newpage
\begin{flushleft}
Acknowledgements
\end{flushleft}
We appreciate enlightening conversations with K.~Machida, 
D.~Agterberg and E.~M.~Forgan. Research at Columbia
 was supported by NSF and NEDO.
 YM and ZQM acknowledge CREST of
the Japan Science and Technology Corporation for its support.

Correspondence and requests for materials  should be addressed to G.M.L.\\ (email: luke@phys.columbia.edu) or Y.J.U. (email: tomo@kirby.phys.columbia.edu).

  \end{document}